\documentclass{emulateapj}
\usepackage{epsfig,graphicx}
\usepackage{natbib}
\usepackage{bm}
\usepackage{natbib}
\usepackage{comment} 
\usepackage{dcolumn}
\shorttitle{Swift J1745$-$26}
\shortauthors{Tetarenko, A.J. et al.}

\submitted{Accepted to ApJ}
\begin{document}

\title{Sub-mm Jet Properties of the X-Ray Binary Swift J1745$-$26}

\author{A.J. Tetarenko\altaffilmark{1}, G.R. Sivakoff\altaffilmark{1}, J.C.A. Miller-Jones\altaffilmark{2}, P.A. Curran\altaffilmark{2}, T.D. Russell\altaffilmark{2}, I.M. Coulson\altaffilmark{3}, \\
S. Heinz\altaffilmark{4}, D. Maitra\altaffilmark{5}, S.B. Markoff\altaffilmark{6}, S. Migliari\altaffilmark{7,8}, G.R. Petitpas\altaffilmark{9}, M.P. Rupen\altaffilmark{10,11}, \\
A.P. Rushton\altaffilmark{12,13}, D.M. Russell\altaffilmark{14}, C.L. Sarazin\altaffilmark{15} }
\altaffiltext{1}{Department of Physics, University of Alberta, CCIS 4-181, Edmonton, AB T6G 2E1, Canada}
\altaffiltext{2}{International Centre for Radio Astronomy Research- Curtin University, GPO Box U1987, Perth, WA 6845, Australia}
\altaffiltext{3}{Joint Astronomy Centre, 660 North A'ohoku Place University Park, Hilo, HI 96720, USA}
\altaffiltext{4}{Astronomy Department, University of Wisconsin-Madison, 475. N. Charter St., Madison, WI 53706, USA}
\altaffiltext{5}{Department of Physics and Astronomy, Wheaton College, Norton, MA 02766, USA}
\altaffiltext{6}{Astronomical Institute `Anton Pannekoek', University of Amsterdam, P.O. Box 94249, 1090 GE Amsterdam, the Netherlands}
\altaffiltext{7}{Department of Astronomy and Meteorology \& Institute of Cosmic Sciences, University of Barcelona, Mart\'i i Franqu\`es 1, 08028 Barcelona, Spain}
\altaffiltext{8}{XMM-Newton Science Operations Centre, ESAC/ESA, PO Box 78, 28691 Villanueva de la Ca\~nada, Madrid, Spain}
\altaffiltext{9}{Harvard-Smithsonian Center for Astrophysics, Cambridge, MA 02138, USA }
\altaffiltext{10}{National Research Council, Herzberg Astronomy and Astrophysics, 717 White Lake Road, PO Box 248, Penticton, British Columbia V2A 6J9, Canada}
\altaffiltext{11}{National Radio Astronomy Observatory, P.O. Box 0, Socorro, NM 87801, USA}
\altaffiltext{12}{Department of Physics, Astrophysics, University of Oxford, Keble Road, Oxford OX1 3RH, UK}
\altaffiltext{13}{School of Physics and Astronomy, University of Southampton, Highfield, Southampton SO17 1BJ, UK}
\altaffiltext{14}{New York University Abu Dhabi, P.O. Box 129188, Abu Dhabi, United Arab Emirates}
\altaffiltext{15}{Department of Astronomy, University of Virginia, P.O. Box 400325, Charlottesville, VA 22904, USA}
\email{tetarenk@ualberta.ca}

\begin{abstract}
We present the results of our observations of the early stages of the 2012--2013 outburst of the transient black hole X-ray binary (BHXRB), Swift J1745$-$26, with the VLA, SMA, and JCMT (SCUBA--2). Our data mark the first multiple-band mm \& sub-mm observations of a BHXRB. During our observations the system was in the hard accretion state producing a steady, compact jet. 
The unique combination of radio and mm/sub-mm data allows us to directly measure the spectral indices in and between the radio and mm/sub-mm regimes, including the first mm/sub-mm spectral index measured for a BHXRB.
Spectral fitting revealed that both the mm (230 GHz) and sub-mm (350 GHz) measurements are consistent with extrapolations of an inverted power-law from contemporaneous radio data (1--30 GHz). This indicates that, as standard jet models predict, a power-law extending up to mm/sub-mm frequencies can adequately describe the spectrum, and suggests that the mechanism driving spectral inversion could be responsible for the high mm/sub-mm fluxes (compared to radio fluxes) observed in outbursting BHXRBs.
While this power-law is also consistent with contemporaneous optical data, the optical data could arise from either jet emission with a jet spectral break frequency of $\nu_{{\rm break}}\gtrsim1\times10^{14}\,{\rm Hz}$ or the combination of jet emission with a lower jet spectral break frequency of $\nu_{{\rm break}}\gtrsim2\times10^{11}\,{\rm Hz}$ and accretion disc emission.
Our analysis solidifies the importance of the mm/sub-mm regime in bridging the crucial gap between radio and IR frequencies in the jet spectrum, and justifies the need to explore this regime further.
\end{abstract}


\maketitle

\section{Introduction}
\setcounter{footnote}{0}
Relativistic jets are powerful, collimated outflows of energy and matter \citep{fen10}. 
While these jets have been studied for decades in accreting sources, the underlying physics that governs jet behaviour is still poorly understood. Despite the many unknowns, it is clear that these jets play a crucial role in the accretion process 
\citep{mei01,merhezdi03,fenbelgal04,falkormar04,galfenkai05,falcke95,fen06,fen10,gal10,cor12,gal14}.

Black hole X-ray binaries (BHXRBs), which contain an accreting stellar mass black hole paired with a companion star \citep{remmc06}, are ideal probes for jet phenomena because they vary on short timescales (days to months). Therefore, many different phases of jet behaviour (jet launching, fading, and quenching) can be analyzed in a single system. 
 Additionally, BHXRBs act as analogues to more observationally inaccessible systems, such as active galactic nuclei (AGN; whose entire outbursts evolve on million-year timescales), where jet feedback is thought to play a key role in galaxy formation and evolution \citep{fab12}.

In BHXRB systems, a compact, steady relativistic jet is present at the beginning of an outburst when the system is typically in the hard accretion state (see \citealt{fenbel04,fenhombel09,bel10} for further discussion on accretion states). 
Jet emission in the hard state reveals a flat to slightly inverted optically thick spectrum ($\alpha\geq0, \textrm{where}\hspace{0.1cm}f_{\nu}\propto\nu^{\alpha}$; \citealt{fen01}) extending from radio through sub-mm frequencies (possibly even up to IR frequencies; \citealt{corfen02,cas10,chat,rus12}). Around the infrared frequencies, the optically thick jet spectrum breaks to an optically thin spectrum ($\alpha<0$; \citealt{rus12}), leading to a rapidly declining flux density with increasing frequency. 

\cite{blandford79} were the first to propose a model for this flat/inverted jet emission, where the overall jet spectrum is generally described as the superposition of individual overlapping synchrotron components originating from different scales along the jet. 
Since then, many variations on this model have been proposed that address and build upon some of the simplifying assumptions of the \cite{blandford79} model \citep[e.g.,][]{hj88,falcke95,mar01,marnowwil05,kai06,peer09,mal13,mal13b,mal14}.  
Although many of these models successfully reproduce the observed jet spectra, they are still limited in describing the main properties and processes within the jet. These detailed jet properties are encoded within the jet spectral energy distribution (SED, e.g., \citealt{heisun03,mar03,heigrimm,marnowwil05,cas09,peer09}). Thus, high quality, well-sampled broadband observations are key in overcoming these challenges.

Given that jet emission is predicted to dominate over other system components (accretion disc and companion star) at frequencies below the near-IR band during the hard state \citep{rus06}, 
the mm/sub-mm regime is crucial to our understanding of the jet.
Currently only a handful of detections of outbursting BHXRBs exist in the mm/sub-mm regime \citep{fenpoo00,fenpoo002,og00,pared00,fender01,rus13,van13,rus14}.
However, recent upgrades to such instruments as the Submillimeter Array (SMA) and the Submillimetre Common User Bolometer Array 2 on the James Clerk Maxwell Telescope (SCUBA--2 on the JCMT), as well as the introduction of new instruments such as the Atacama Large Millimetre Array (ALMA) are enabling more detections of these sources with flux densities on the order of a mJy (or even as low as tens of $\mu{\rm Jy}$ with ALMA).

With mm/sub-mm data we are able to fill in a gap of $\sim2$ orders of magnitude in frequency in our broadband coverage. This is especially important when attempting to constrain the location of the spectral break, which is postulated to mark the location where particles are first accelerated to a power-law distribution in the jet \citep{mar01,marnowwil05,pol10,pol13,pol14}. Additionally, the flux and frequency of this spectral break can reveal insights into universal jet properties, such as minimum radiative jet power, key system parameters, such as accretion rate, black hole mass, radius of the inner accretion disc and magnetic field strength \citep{heisun03,mar03,heigrimm,marnowwil05,cas09,peer09,chat,pmark,rus13}, as well as uncover physical conditions in the jet,  
such as the base jet radius, velocity, and opening angle (\citealt{rus13,rus12}; \citealt{rus14}). 
However, the break has only been directly detected in three black hole sources, GX 339--4 \citep{cor02,gan11}, MAXI J1659$-$152 \citep{van13}, and V 404 Cyg (additionally in this source the radio--IR spectrum was curved, requiring a second break in the flat/inverted spectral regime; \citealt{rus12}), and indirectly constrained\footnote{Observing the spectral break indirectly refers to estimating the spectral break frequency by interpolating between the radio and IR-optical power-laws (through the unknown mm/sub-mm regime) as opposed to directly observing the spectral break within the data.} in other black hole sources. Most recently \cite{rus12} presented indirect jet break constraints for a large sample of sources, and similar results are available for Cyg X$-$1 \citep{rah11,now05}, and MAXI J1836$-$194 (\citealt{rus13a}; \citealt{rus14}). This indirect interpolation process introduces significant uncertainties in the derived location of the break (up to an order of magnitude in MAXI J1836-194; \citealt{rus13a}) and requires that the radio--sub-mm spectrum can be accurately represented by a single power-law. Data in the mm/sub-mm part of the spectrum allow us to make direct spectral measurements intermediate between radio frequencies and the spectral break, mitigating the uncertainties that come with interpolation and testing this single radio--sub-mm power-law assumption.

Recent results from observing campaigns of the BHXRB sources MAXI J1836--194 and MAXI J1659--152 show evolving SEDs, suggesting an evolving jet break that appears to tend toward lower radio frequencies as the accretion rate increases, and the compact jet begins to switch off during the transition to softer states at the peak of the outburst \citep{rus13a,van13,rus14}. Additionally, \cite{cor13} found that the jet break in GX 339--4 evolved as the jets switched back on in the reverse state transition. With mm/sub-mm data we can directly track the evolution of the break through mm/sub-mm frequencies and down to the radio band.
Tracking the break could allow us to correlate the changing break frequency with accretion properties, such as X-ray hardness (\citealt{rus14} find tentative evidence that the break frequency may correlate with X-ray hardness in MAXI J1836--194), which is essential in understanding what physical processes  
are driving changes within the jet. 

Although mm/sub-mm observations of BHXRBs are sparse, the few mm/sub-mm detections of BHXRBs to date in the literature (e.g., \citealt{fender01,rus13}) have measured considerably higher flux densities than seen at radio frequencies ($\sim40-70\, \textrm{mJy}$).  
These high mm/sub-mm fluxes could be ``anomalous", in which the excess emission (above that of a flat spectrum extending across radio frequencies) at mm/sub-mm frequencies was produced by a yet unknown process not included in standard jet models (e.g., see \citealt{mar01}). On the other hand, high mm/sub-mm fluxes could be the result of a more inverted (rather than flat) radio through sub-mm spectrum (e.g., inverted radio through IR spectra have been observed in V404 Cyg; \citealt{galc05,hyn05}, A0620--00; \citealt{gala07}, and XTE J1118$+$480; \citealt{fender01}).
Therefore, it is essential to first understand the origin of the mm/sub-mm flux in BHXRBs before dynamic broad-band SEDs can be used to constrain jet properties. 

 \subsection{Swift J1745$-$26}
Swift J174510.8-262411 (also known as Swift J1745$-$26) is a transient black hole candidate \citep{vov12} source discovered in the Galactic centre region ($l=2.11^\circ, b=1.40^\circ$) by NASA-led Swift Burst Alert Telescope (BAT; $15-50{\rm \, keV}$) on 2012 September 16 \citep{cum12a}. An X-ray counterpart was confirmed in the hard X-rays ($0.2-10{\rm \, keV}$) by the X-ray telescope (XRT) on the Swift satellite on 2012 September 17 \citep{cum12b,shar12}. 
X-ray spectral and timing observations from Swift and the International Gamma-Ray Astrophysics Laboratory (INTEGRAL) were used to classify this source as a low mass X-ray binary (LMXB)
black hole candidate system \citep{white95}. In addition, this outburst was classified as ``failed" \citep{bel12,sbar13}, as it did not reach the soft state (see \citealt{brok04} for further discussion on failed outbursts). A radio detection was made on 2012 September 17--18 with the Karl G. Jansky Very Large Array (VLA) in the 5.0 and 7.45 GHz bands of $6.8\pm0.1\, \textrm{mJy}$ and $6.2\pm0.1\, \textrm{mJy}$, respectively, suggesting that the emission likely originated from a partially self-absorbed compact jet ($\alpha=-0.22\pm0.09$; \citealt{mill12}). 
Follow-up radio observations with the Australia Telescope Compact Array (ATCA) confirmed the presence of a partially self-absorbed compact jet ($\alpha=-0.05\pm0.04$; \citealt{cor12a}). Further, \cite{curr14} performed a detailed radio frequency study of the entire outburst and \cite{kale14} analyzed the decay of the outburst at X-ray, optical, and radio frequencies. The source outburst ended approximately 2013 June 20, when it could no longer be detected by Swift BAT.

Between 2012 September 20 and 2012 September 26 (i.e., early in the outburst), we obtained quasi-simultaneous radio and multiple band mm \& sub-mm observations of the source, combining data from the VLA, SMA, and JCMT SCUBA--2. 
 These data afford us the unique opportunity to directly measure the spectral indices in and between the radio and mm/sub-mm regimes. 
  In \S 2 we describe the data collection and reduction processes for the SMA, JCMT SCUBA--2, and VLA. In \S 3 we present the radio through sub-mm spectrum, outline the spectral fitting process, and show the results of the spectral fits. \S 4 contains an interpretation of the spectral behaviour presented in \S 3, as well as a discussion on the origin of high mm/sub-mm fluxes and variability at radio frequencies. A summary of the results is presented in \S 5.

\section{Observations and Data Analysis}
\subsection{SMA}
SMA observations (Project Code: 2012A-S055) of Swift J1745$-$26 were taken on three nights at 230 GHz (see Table~\ref{table:imflu}).
The very extended array configuration was used with a total of 5 antennas on September 20th, and 7 antennas on September 22nd and 25th (out of a possible 8 antennas). All observations were made in double bandwidth mode (single receiver, 4 GHz bandwidth) and with precipitable water vapour measurements of $\sim4\,{\rm mm},\, 3\,{\rm mm},\, {\rm and}\, 1\,{\rm mm}$ on September 20, 22, and 25 respectively. We used 1924-292 and 3C84 as bandpass calibrators, 1924-292 and NRAO 530 as gain calibrators, and Neptune as a flux calibrator\footnote{The SMA calibrator list can be found at http://sma1.sma.hawaii.edu/callist/callist.html.}. Data were reduced in the Common Astronomy Software Application (CASA; \citealt{mc07}) using standard procedures outlined in the Casaguides for SMA data reduction. Currently, CASA is unable to handle SMA data in its original format;  therefore the SMA scripts, \textit{sma2casa.py} and \textit{smaImportFix.py} were used to convert the data into CASA MS format and perform the $T_{{\rm sys}}$ correction\footnote{Links to the SMA Casaguides and these scripts are publicly available at www.cfa.harvard.edu/sma/casa.}. On September 20 poor weather conditions, the limited number of antennas, and phase de-correlation led to overall poor quality data that prevented us from placing any constraints on source brightness. 
Swift J1745$-$26 was significantly detected on both the 22nd and 25th, and careful phase self-calibration, ensuring we obtained smoothly varying solutions with time, was used to correct for any phase de-correlation that occurred on these nights.  
All Swift J1745$-$26 flux densities, as measured by fitting a point source in the image plane with imfit in CASA, are presented in Table~\ref{table:imflu}\footnote{Note that difmap was also used to model fit the visibilities in the uv-plane rather than the image plane. All flux density values obtained from fitting in the uv-plane were consistent with those found from fitting in the image plane.}.

\subsection{JCMT SCUBA--2}
JCMT observations (Project Code: M12BC25) of Swift J1745-26 were taken on the night of 2012 September 21 in the $850\mu\rm{m}$ band (352.697 GHz). Two observations spaced approximately 45 minutes apart were obtained with the SCUBA--2 detector \citep{chap,holl}. Observations of the flux calibrators \citep{demp} Uranus, CRL618, and CRL2688 were present but we chose to use CRL2688 as it was closest in both time and space to our target source. The daisy configuration was used, producing a 3 arcmin map. During the observations we were in the Grade 2 weather band with a 225 GHz opacity of 0.07-0.08. Data were reduced in the Starlink package using standard procedures outlined in the SCUBA--2 cookbook and the SCUBA--2 Quick Guide\footnote{www.jach.hawaii.edu/JCMT/continuum/scuba2/ \\
scuba2\_quickguide.html}.
We had $\sim30$ minutes on source per observation for a total time of $\sim60$ minutes on source.  
The Swift J1745$-$26 flux density values calculated for each observation as well as the co-added observations are listed in Table~\ref{table:imflu}, where the uncertainty in the flux density measurements include a substantial contribution from the uncertainty in the flux conversion factor (FCF).  Note that data in the $450\mu\rm{m}$ (666.205GHz) band were obtained simultaneously with the $850\mu\rm{m}$ band, but the source was not significantly detected at $450\mu{\rm m}$, with an upper limit at $450\mu{\rm m}$ that was not strongly constraining ($\sim250\,{\rm mJy}$).

\subsection{VLA}
We re-reduced the VLA data presented in \cite{curr14} during the epochs taken within two days of our mm/sub-mm observations (see Table~\ref{table:imflu}). These radio epochs are quasi-simultaneous with our mm/sub-mm epochs, and thus represent the best opportunity to constrain the jet spectrum. The re-reduction was performed in CASA following the same flagging, calibration, and imaging procedure as \cite{curr14}, with more up-to-date antenna positions that were not published at the time of the initial analysis. All flux densities presented include the conventional VLA systematic errors of 1\% ($< 10{\rm GHz}$), 3\% ($10-40{\rm GHz}$), and 5\% ($> 40{\rm GHz}$).
Poor weather on September 26 meant that observations above 32 GHz suffered from complete phase de-correlation. As such, these measurements may include significant, unaccounted for errors and were not used in our analysis (further discussion is presented in \S 3.1).

\renewcommand\tabcolsep{2.5pt}
\begin{table}[t!]
\small
\caption{Flux Densities of Swift J1745$-$26 for Radio, mm, and sub-mm Frequency Data from the VLA, SMA, and JCMT}\quad
\centering
\begin{tabular}{ cccccc }
 \hline\hline
   &{\bf Date} &{\bf MJD}& {\bf Time on}&{\bf Freq.}&{ \bf Flux}\\
  & {\bf (2012)}&&\bf{Source}&{\bf (GHz)}&{\bf (mJy)}  \\
   &&&{\bf (min)}&&\\[0.05cm]
  \hline
VLA& Sep 20&56190.09&10.97 &5.0\phn&17.72$\pm$0.25\phantom{\footnotemark[3]}\\[0.05cm]
VLA& Sep 20&56190.09&10.97 &7.5\phn&17.97$\pm$0.22\phantom{\footnotemark[3]}\\[0.05cm]
SMA& Sep 20&56190.12&186.6 &219.2&\dots\footnotemark[1]\\[0.05cm]
SMA& Sep 20&56190.12&186.6 &232.6&\dots\footnotemark[1]\\[0.05cm]
JCMT& Sep 21 &56191.27&33.6 &352.7&45.27$\pm$7.16\\[0.05cm]
JCMT& Sep 21&56191.29&33.6 &352.7&37.23$\pm$7.12\\[0.05cm]
JCMT& Sep 21 &\phn co-added\footnotemark[2]&69.6 &352.7&39.85$\pm$5.04\\[0.05cm]
SMA& Sep 22&56192.13&160.0 &219.2&32.10$\pm$1.30\\[0.05cm]
SMA& Sep 22&56192.13&160.0&232.6&35.32$\pm$1.71\\[0.05cm]
VLA& Sep 23&56193.07&1.99 &5.0\phn&25.83$\pm$0.35\\[0.05cm]
VLA& Sep 23&56193.07&1.99 &7.5\phn&26.19$\pm$0.36\\[0.05cm]
VLA& Sep 23&56193.06&1.99 &20.8&26.78$\pm$1.68\\[0.05cm]
VLA& Sep 23&56193.06&1.99 &25.9&26.73$\pm$1.67\\[0.05cm]
VLA& Sep 25&56195.03&2.49 &5.0\phn&25.82$\pm$0.35\\[0.05cm]
VLA& Sep 25&56195.03&2.49 &7.5\phn&25.26$\pm$0.41\\[0.05cm]
VLA& Sep 25&56195.02&3.99 &20.8&28.35$\pm$1.11\\[0.05cm]
VLA& Sep 25&56195.02&3.99 &25.9&27.90$\pm$1.11\\[0.05cm]
SMA& Sep 25&56195.13&164.4 &219.2&35.11$\pm$1.11\\[0.05cm]
SMA& Sep 25&56195.13&164.4 &232.6&37.83$\pm$1.52\\[0.05cm] 
VLA& Sep 26&56196.03&2.49 &1.4\phn&21.83$\pm$0.62\\[0.05cm]
VLA& Sep 26&56196.03&2.49 &1.8\phn&23.58$\pm$0.57\\[0.05cm]
VLA& Sep 26&56196.02&3.99 &31.5&31.31$\pm$1.03\\[0.05cm]
VLA& Sep 26&56196.02&3.99 &37.5&36.46$\pm$1.67\\[0.05cm]
VLA& Sep 26&56196.01&3.99 &41.5&38.86$\pm$3.33\\[0.05cm]
VLA& Sep 26&56196.01&3.99 &47.5&40.98$\pm$3.52\\[0.1cm] \hline

\end{tabular}
\footnotetext{No flux density values are given for 2012 September 20 as poor quality data prevented us from constraining the source brightness in this epoch.}
\footnotetext{Co-added images were combined using a pixel-by-pixel variance weighting technique implemented by the \textit{wcsmosaic} task in Starlinks KAPPA package.}
\label{table:imflu}
\end{table}
\renewcommand\tabcolsep{6.0pt}

\section{Results}
\subsection{Light Curves and Spectrum}
\begin{figure}[t!]
\centering
 {\includegraphics[width=9.0cm,height=7.0cm]{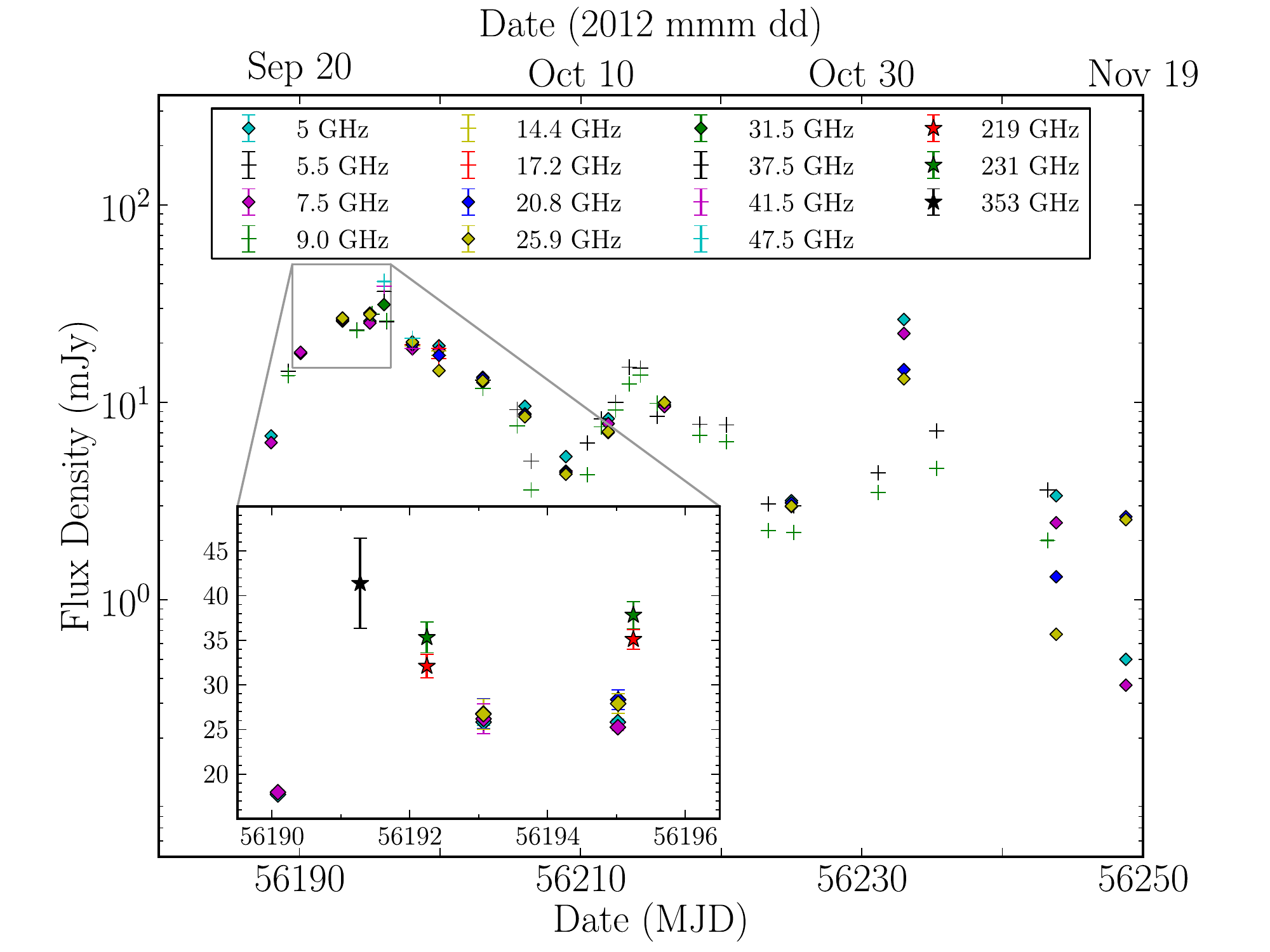}}
\caption[Radio and mm/sub-mm Light Curves of Swift J1745$-$26]{\label{fig:lcurv}Light curves of Swift J1745$-$26 during the first 60 days of its 2012--2013 outburst. Main panel: Radio (VLA and ATCA) light curves spanning the entire outburst, taken from \cite{curr14}. Inset Panel: Radio (VLA; diamonds) and mm/sub-mm (SMA/JCMT; stars) light curves during the hard state of the outburst; this panel contains only the data analyzed in this paper. Flux densities in the 1.5, 31.5, 37.5, 41.5, 47.5 GHz radio bands are not shown in the inset panel for clarity.
}
\end{figure}

Figure~\ref{fig:lcurv} shows the radio frequency light curve spanning the first 60 days of the 2012-2013 outburst of Swift J1745$-$26 \citep{curr14}, with an inset panel that zooms in on the 6-day period at the beginning of the outburst (September 20--26) for which we have mm/sub-mm coverage.

Our observations show that, at mm/sub-mm frequencies, the source may have evolved somewhat differently than at radio frequencies, possibly even not evolving at all in the mm/sub-mm bands.
 At radio frequencies we see a rise in source brightness, culminating in a brightness peak at 2012 September 26 (corresponding to the peak observed radio brightness over the outburst; \citealt{curr14}). Whereas, at mm/sub-mm frequencies, while the source brightness appears to remain relatively constant over our observations, poor temporal sampling prevents us from ruling out any variability, 
or drawing any further conclusions on differences in temporal behaviour between the radio and mm/sub-mm regimes.

 The radio--sub-mm spectrum of Swift J1745$-$26 through different epochs during its 2012--2013 outburst can be seen in Figure~\ref{fig:specfitind} (along with power-law fits, described in \S 3.2). All of the radio and mm/sub-mm flux measurements were made within $\lesssim2$
days (i.e., quasi-simultaneous) of each other in an attempt to best constrain the jet spectrum. 
Clear flux variability is observed between epochs at radio frequencies. Additionally, a more inverted spectrum is seen at higher radio frequencies in the 2012 September 26 epoch, where we see an apparent flux increase of $\sim15\;\textrm{mJy}$ between $\sim30-50\,\textrm{GHz}$ (transparent green points on Figure~\ref{fig:specfitind}). 
However, due to complete phase de-correlation, attempts at phase self-calibration produced solutions that are not smoothly varying with time, but look like pure noise above 32 GHz (see \S 2.2 above). Therefore, while the upward trend in the highest frequency data from 2012 September 26 is intriguing, it is possibly (and perhaps most likely) an artifact of the self-calibration process in the presence of unstable atmospheric conditions. We caution against over-interpreting this feature and the remainder of our analysis will not include these data points.

The source brightness at mm/sub-mm frequencies is higher than that at radio frequencies (up to a factor of $2$), similar to other BHXRB sources with mm/sub-mm detections \citep{fenpoo00,fenpoo002,og00,fender01,rus13a,van13,rus14}. We discuss possible causes of high mm/sub-mm fluxes further in \S4.1 and \S4.3 below.

\subsection{Spectral Fitting in the Individual (radio \& mm/sub-mm) and Global (radio through sub-mm) Regimes}
\begin{figure}[t!]
\centering
 {\includegraphics[width=9cm,height=7cm]{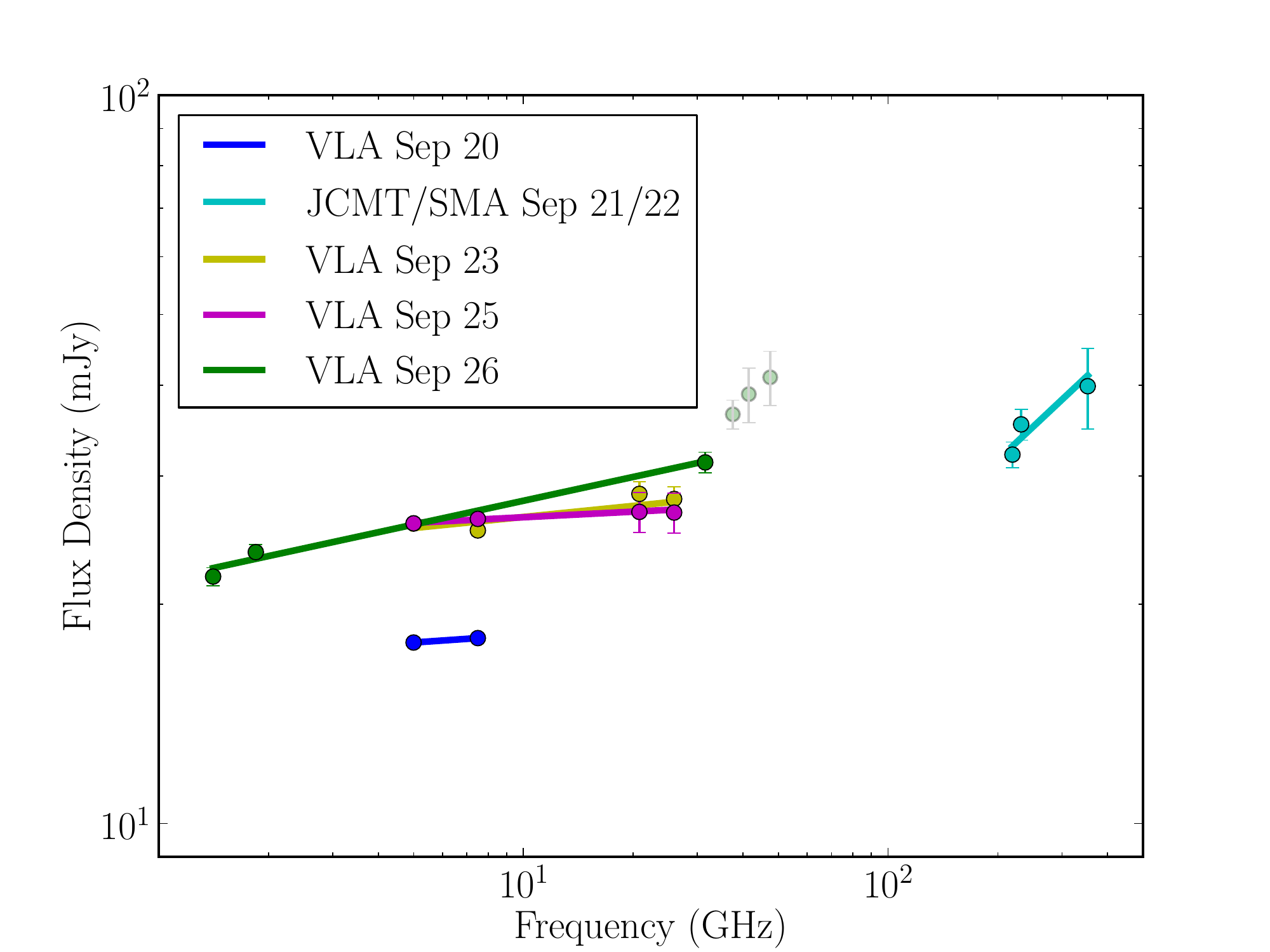}}
\caption[Radio and mm/sub-mm Spectral Fits of Swift J1745$-$26]{\label{fig:specfitind}Radio and mm/sub-mm spectra for various epochs during the hard state of the 2012--2013 outburst of Swift J1745$-$26. The solid lines indicate the power-law fits to the data (Table~\ref{table:sp}).  The transparent green points indicate measurements that may suffer significant, unaccounted for, errors (see \S 2.2), they are not included in the fits and are presented here for comparison purposes only. The SMA measurements from 2012 Sep 25 are not included in the mm/sub-mm index fit as there is a $\sim4$ day separation between these data and the JCMT 350 GHz data. The mm/sub-mm index is more inverted compared to the radio indices across all epochs, although this result has a high uncertainty.}
\end{figure}

\begin{figure}[t!]
\centering
 {\includegraphics[width=9cm,height=7cm]{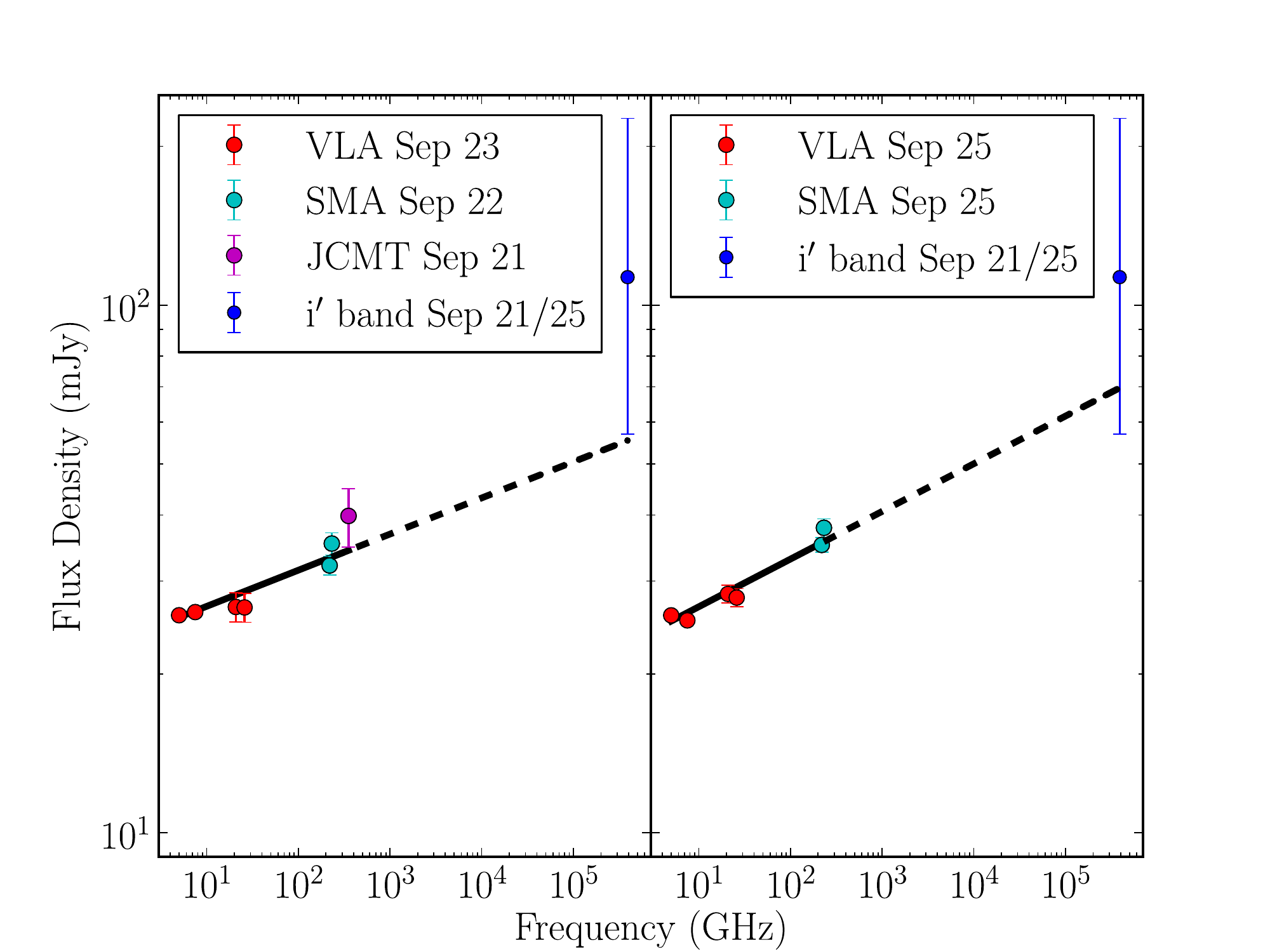}}
\caption[Radio through sub-mm Global Spectral Fits of Swift J1745$-$26]{\label{fig:specfitglob}Radio through sub-mm spectral fits during the hard state of the 2012--2013 outburst of Swift J1745$-$26. At different epochs, the solid black lines indicate the power-law fits to the data (Table~\ref{table:sp}).  The dotted black line indicates an extrapolation of the radio--sub-mm power-law to the optical regime,  where the optical data point (from combined i' band measurements on 2012 Sep 21 and 25; \citealt{mun13}) is not included in the fit. A power-law is clearly well representative of the radio through sub-mm data at these times, and its interpolation is consistent with the measured optical point as well (see \S 3.3 and \S 4.2 for further discussion).}
\end{figure}

\begin{figure}[t!]
\centering
 {\includegraphics[width=9cm,height=7cm]{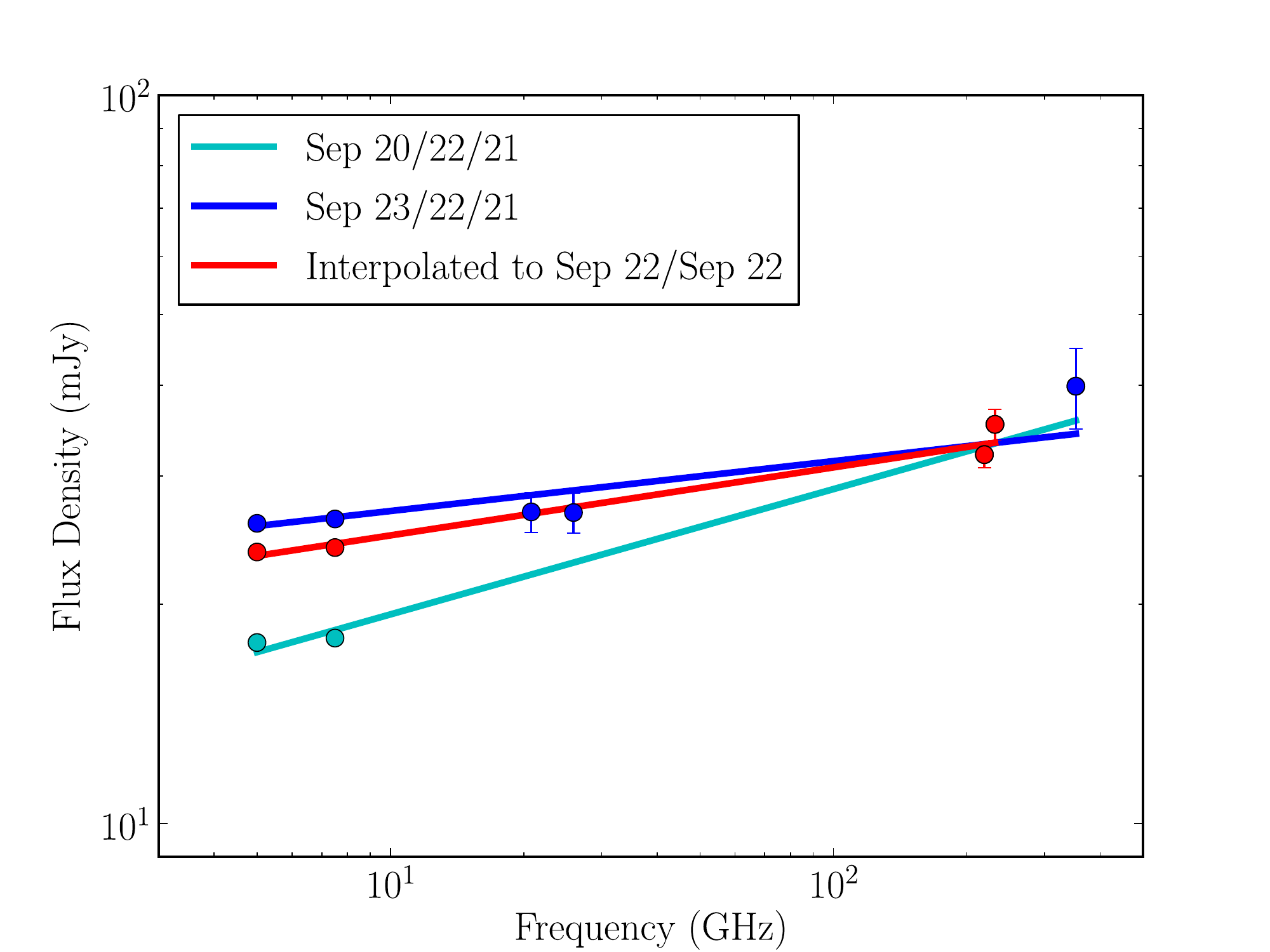}}
\caption[Interpolated Global Spectral Fits of Swift J1745$-$26]{\label{fig:specint}Radio through sub-mm spectral fits for various epochs during the hard state of the 2012--2013 outburst of Swift J1745$-$26. The solid lines indicate the power-law fits to the spectra. The legend indicates the epochs of data included in each fit in order of increasing frequency (i.e., radio/mm/sub-mm).  All spectral indices complete with errors can be seen in Table~\ref{table:sp}. The combination of quasi-simultaneous data (separated by up to 2 days) and flux variability between epochs clearly effects the results of our spectral fitting and thus our interpretation of the jet spectrum.}
\end{figure}

\begin{figure}[t!]
\centering
 {\includegraphics[width=9cm,height=7cm]{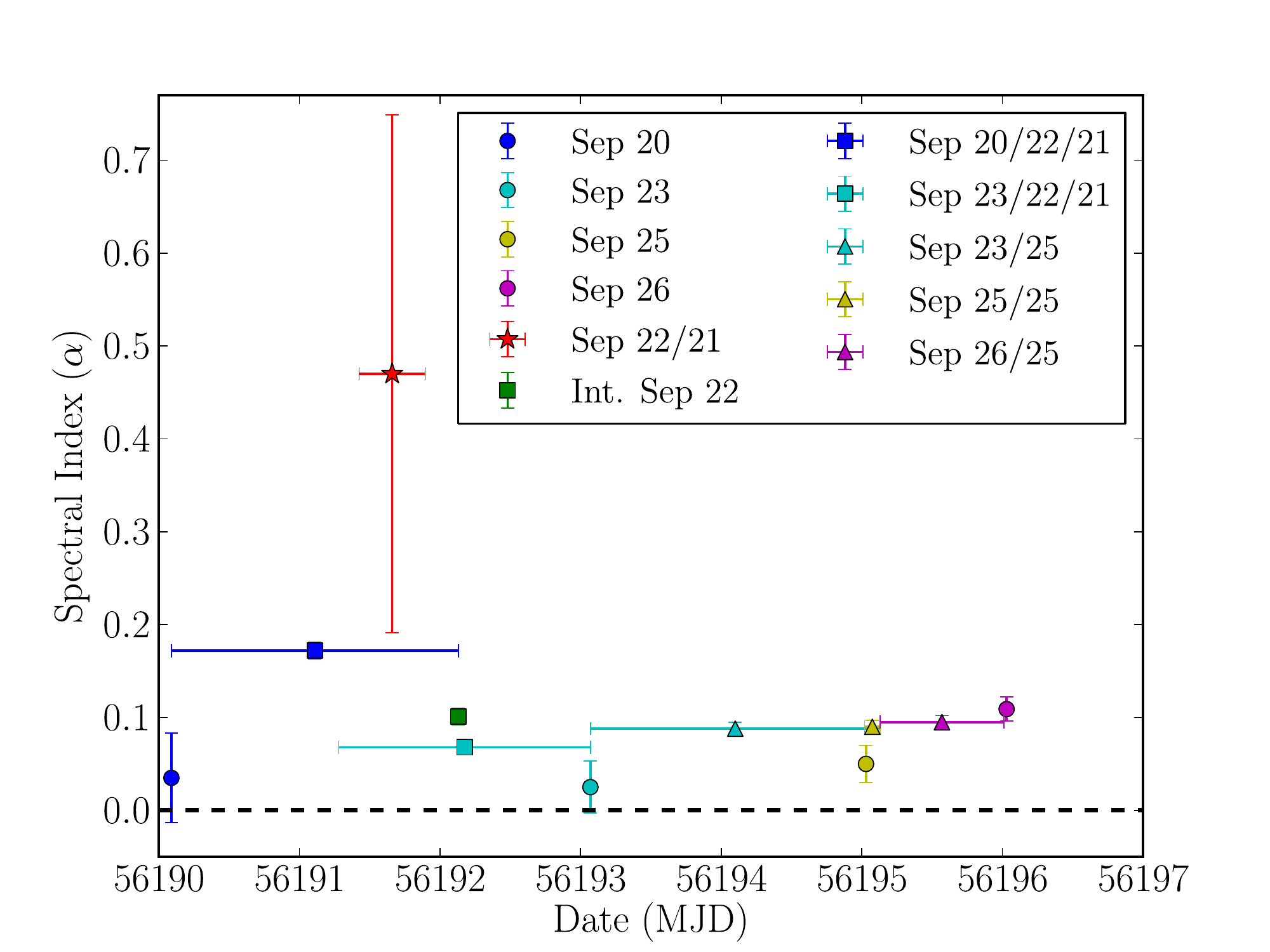}}
\caption[]{\label{fig:spes}Radio and mm/sub-mm spectral indices of Swift J1745$-$26. Circles indicate epochs containing only radio data, stars indicate epochs containing only mm/sub-mm data, squares indicate epochs containing radio, mm and sub-mm data, and triangles indicate epochs containing radio and mm data. For epochs that contain quasi-simultaneous data sets the midpoint is plotted and horizontal error bars indicate the entire time range. The black dashed line indicates a flat spectral index; $\alpha=0.0$. All spectral index measurements plotted here can be seen in Table~\ref{table:sp}.  Data sets containing mm/sub-mm data show more inverted spectral indices than those containing only radio data.}
\end{figure}

 Standard jet models \citep{blandford79} predict that a self-absorbed compact jet would produce emission that follows a single power-law from radio through sub-mm frequencies, and as such we would expect $\alpha_{\tiny{\textrm{radio}}}\sim\alpha_{\tiny{\textrm{mm/sub-mm}}}$. On the other hand, while this simple jet model has been proven to match observations at radio frequencies in multiple sources, its predictions have not been tested at higher mm/sub-mm frequencies. To test these models
in the mm/sub-mm regime, we chose to fit a power-law to our jet spectrum, using an 
affine-invariant ensemble sampler for Markov chain Monte Carlo \citep[MCMC][]{for2013} and 
the standard least squares algorithm in logarithmic space for power-law fitting to determine the best-fit and obtain accurate (1-$\sigma$) parameter uncertainties.
 Figure~\ref{fig:specfitind}, Figure~\ref{fig:specfitglob}, and Figure~\ref{fig:specint} display various spectral fits in the individual (here a power-law was fit across radio only or mm/sub-mm only data) and global (here a constant slope power-law extending from radio through sub-mm frequencies was fit to the data) regimes for different epochs of data, Figure~\ref{fig:spes} displays all spectral indices calculated over the time period we had data, and Table~\ref{table:sp} displays the results of all the spectral fits.   
 
In the individual regime fits (Figure~\ref{fig:specfitind}), both the radio and mm/sub-mm are fit reasonably well with a single power-law (low $\chi^2$ given the number of degrees of freedom). 
All of the epochs of radio data display a relatively flat spectral index ($\alpha\sim0.0-0.1$) as expected from a simple self-absorbed jet.
In the mm/sub-mm regime, the spectrum is more inverted ($\alpha=0.5\pm0.3$) than at radio frequencies in both epochs, albeit this is only a $1.5\sigma$ result. The high level of uncertainty in this index is mainly due to the poorly constrained 350 GHz data point, in which limited time on source ($\sim1$ hr) led to weak limits on flux density. 
As a result, all spectral indices (radio and mm/sub-mm in all epochs) are consistent with each other at the $3\sigma$ confidence level.

In the global regime, we once again see that the data are reasonably well fit by a power-law (e.g., Figure~\ref{fig:specfitglob}). However, it is clear that some epochs are fit better with a power-law than others (with poorer fits showing deviations from a single power-law at the $\sim2\sigma$ level). In particular, the data sets containing the September 20 radio data show a much poorer fit than those containing the September 23 radio data, even when both data sets are paired with the same mm/sub-mm data (see Table~\ref{table:sp}). We believe that these poorer quality fits mainly result from flux variability at radio frequencies occurring between the days on which we have data, as the radio flux density was stable from September 23--25, whereas it rose significantly between September 20 and 23.

In an effort to mitigate this effect, we interpolated the radio frequency data to the days on which the SMA mm data were taken (see Fig.~\ref{fig:specint} and Table~\ref{table:sp}).

\begin{table*}
\small
\renewcommand\tabcolsep{4.0pt}
\caption{Spectral Indices for Individual and Global Epochs During the 2012--2013 
Outburst of Swift J1745$-$26}\quad
\centering
\begin{tabular}{lcccccc}
\hline\hline
{\bf Frequency}&{ \bf Figure}&{\bf Data Sets}&{\bf Power-Law}&${\bm \chi^2}$&{\bf dof\,}\footnotemark[1]&${\bm P_{{\rm {\bf null}}}}$\footnotemark[2]\\
{\bf Band(s)}&{\bf (Color/Panel)}&{\bf Fitted (2012 Sep dd)}&{\bf Spectral Index (${\bm \alpha}$)}&&&\\[0.05cm]
\hline
radio&\ref{fig:specfitind} (blue)&20&$0.035\pm0.048$\footnotemark[3]&0&0&1.00\\[0.05cm]
radio&\ref{fig:specfitind} (magenta)&23&$0.025\pm0.028$\phn&0.06&2&0.97 \\[0.05cm]
radio&\ref{fig:specfitind} (yellow)&25&$0.050\pm0.020$\phn&4.98&2&0.08 \\[0.05cm]
radio&\ref{fig:specfitind} (green)&26&$0.109\pm0.013$\phn&1.68&1&0.19\\[0.05cm]
mm+sub-mm&\ref{fig:specfitind} (cyan)&22+21&$0.470\pm0.279$\phn&1.30&1&0.25 \\[0.05cm]
radio+mm+sub-mm&\ref{fig:specint}(cyan)&20+22+21&$0.172\pm0.009$\phn&11.56&3&0.01\\[0.05cm]
radio+mm+sub-mm&\ref{fig:specint}(blue) \& \ref{fig:specfitglob}(left)&23+22+21&$0.068\pm0.008$\phn&\phn6.06&5&0.30\\[0.05cm]
radio+mm&\dots&23+25&$0.088\pm0.007$\phn&\phn7.39&4&0.12\\[0.05cm]
radio+mm&\ref{fig:specfitglob}(right)&25+25&$0.090\pm0.007$\phn&11.37&4&0.02\\[0.05cm]
radio+mm&\dots&26+25&$0.095\pm0.007$\phn&\phn4.97&3&0.17\\[0.05cm]
radio+mm&\dots&25/26+25&$0.092\pm0.006$\phn&15.57&7&0.03\\[0.05cm]
radio+mm+sub-mm&\ref{fig:specint}&interp. to 22&$0.101\pm0.009$\phn&\phn3.14&2&0.21\\[0.1cm]
\hline
\end{tabular}
\footnotetext{Degrees of freedom}
\footnotetext{Null hypothesis probability, where the null hypothesis is that the data is perfectly represented by a power-law.}
\footnotetext{Note the spectral index calculated with 2012 September 20 radio data has a higher level of uncertainty than the other indices as it is calculated using only two flux density/frequency measurements.}
\label{table:sp}
\end{table*}
\renewcommand\tabcolsep{6.0pt}

While the quality of the fits improved following interpolation, we still noticed some epochs with radio and mm data sets separated by only hours show poorer quality fits than those epochs with radio and mm data separated by days (for instance, the data sets that both contain SMA mm data from September 25, but differ in the fact that they contain VLA radio data from September 23 or September 25; see Table~\ref{table:sp}).
As such we opted to take a closer look at the individual radio bands by examining data on a per sub-band basis rather than a per baseband basis. 

We fit a power-law within the lower frequency radio bands using per sub-band based data (in higher frequency radio bands, $>26\,{\rm GHz}$, the fractional bandwidth, $\frac{\Delta\nu}{\nu}$, is so narrow that it is not particularly useful to measure flux densities on a per sub-band basis).  While all the bands follow a single power-law quite accurately as expected, surprisingly the indices within the bands do not always match the global indices across the bands (with deviations between $2-3\sigma$). 
As VLA data are observed sequentially (not simultaneously) in some frequencies (e.g., while 5 and 7.5 GHz are observed simultaneously, as are 21 and 26 GHz, 5 and 21 GHz are observed sequentially), this result could be suggestive of rapid variability occurring on even shorter timescales, perhaps less than our observational cadence (minutes rather than days). 
We discuss possible causes of the potential rapid variability further in \S4.4 below.

\subsection{Jet Spectral Break Constraints}
 \cite{mun13} present optical frequency measurements (Sloan i', Bessel I, V \& R Johnson) of Swift J1745$-$26 that are coincident with our mm/sub-mm observations.
 To calculate de-reddened flux densities we use the prescription in \cite {card89}, adopting the value of $n_{{\rm H}}=(2.18\pm0.25)\times10^{22}\,{\rm cm}^{-2}$ from \citet{kale14}, which corresponds to $A_v=9.86\pm1.20$ \citep{guv09}. This leads to large uncertainties on the de-reddened fluxes, especially for the (bluer) optical bands, that prevents us from constraining the normalization and spectral index in the optical/IR regime. Therefore, we cannot estimate the frequency of the jet spectral break through interpolation between the radio/sub-mm and optical/IR power-laws, as is typically done.
 
We can compare the de-reddened flux density in the i' band ($3.93\times10^{14}\,{\rm Hz}$) from the 2m Faulkes Telescope South (FTS) of $f_{i'}=113^{+113}_{-\phn56}\,{\rm mJy}$ (corresponding to the average of all i' band observations, i' = $17.7\pm0.1$) to contemporaneous radio and mm/sub-mm measurements. We find that the i' flux density lies above the extrapolated radio--sub-mm spectrum in all our global fits (see Figure~\ref{fig:specfitglob}); however, the error of the i' flux density allows for it to be consistent or below the extrapolated radio--sub-mm flux densities. Since high extinction prevented \cite{mun13} from obtaining reliable optical/IR spectral information needed to place accurate constraints on the optical emission processes in Swift J1745-26, we consider two extreme cases for the origin of the i'flux density, emission arising completely from the jet or minimal emission arising from the jet. In the latter case, the reprocessing of X-rays in the outer accretion disc \citep[e.g.,][]{van98,char06} is the dominant emission process \citep{rus06}.

If the i' flux density arises entirely from jet emission, than the jet spectral break could be located either near or bluer than the i' band.  To better estimate the lower limit for the jet spectral break, we performed Monte Carlo simulations that include extrapolations to i'-band of our radio to sub-mm power-law fit (and its errors) and the fully propagated errors on the i' flux density. When the simulated i' flux densities were above the power-law extrapolation, we assigned a lower limit of $3.93\times10^{14}\,{\rm Hz}$. When the simulated i' flux densities were below the power-law extrapolation, we assumed a canonical optically thin spectral index of $\alpha=-0.7$ \citep{mirrod99} and solved for the jet spectral break frequency. Our 99\% confidence interval lower limits for the jet spectral break frequency are $1.3\times10^{14}\,{\rm Hz}$ and $1.0\times10^{14}\,{\rm Hz}$ for data near September 21 and September 25, respectively.

If the i' flux density arises (almost) entirely from accretion disc processes, then that implies a significantly lower jet spectral break frequency. This assumption is not unreasonable given the high accretion disc fractions of emission in i' band in other BHXRBs and the fact that typically in a BHXRB the radio and optical flux densities are about the same to within an order of magnitude \citep{rus06}. Thus, in this case, we can only constrain a lower limit on the jet spectral break frequency using the radio through sub-mm data. Given the small errors on our mm-data from the SMA and the fact that they are always above the extrapolation of the radio spectral index, this sets a lower limit to the jet spectral break at the upper side band frequency for the SMA data, $\nu_{\rm break} \gtrsim 2.3\times10^{11}\,{\rm Hz}$. We note that if we use the same assumptions as in the above paragraph, then only $\sim 0.03\%$ of the i' flux density needs to arise from jet emission to be consistent with this jet spectral break frequency.

\section{Discussion}

\subsection{Interpretation of Spectral Indices}

The results of our spectral fitting in the individual regimes show that the radio spectral indices are all nearly flat ($\alpha\sim0-0.1$) as expected for a typical partially self-absorbed synchrotron jet  \citep{fen01}, and similar to what has been seen in other BHXRB sources in the hard state (\citealt{rus12} and references therein; \citealt{rus13a}; \citealt{van13}). 
Conversely, the mm/sub-mm index appears more inverted when compared to all the radio indices.  

We have to be cautious when taking this result at face value as the highly inverted mm/sub-mm index is poorly constrained due to weak limits on the flux measurement at 350 GHz. Therefore, while the data suggest conflicting spectral behaviour between radio and mm/sub-mm regimes (and in turn the possibility of changing physical conditions in the jet), the plausibility of systematic problems with the relative flux measurements (i.e., spectral indices), combined with the fact that all indices are consistent within $3\sigma$, suggests that the discrepancy between regimes could be entirely due to the combination of systematic and statistical uncertainties. 

If we compare the indices calculated with only radio measurements to the global indices containing the same radio measurements paired with the mm/sub-mm measurements (see Figure~\ref{fig:spes}), the global indices are noticeably more inverted (driven higher by the bright mm/sub-mm measurements). For instance,
there is a $\sim2.5\sigma$ difference between the 2012 September 23 radio spectral index (Figure 2 yellow) and the radio through sub-mm index from data interpolated to 2012 September 22 (Figure 3 red). 
Therefore, when comparing absolute fluxes between the radio and mm/sub-mm regimes rather than relative fluxes (compare radio only and mm/sub-mm only indices), we find even stronger evidence for spectral inversion driving high mm/sub-mm fluxes in BHXRBs.

Recent work \citep{kai06,peer09} suggests that a more inverted jet spectrum (when compared to a flat, $\alpha\sim0$ spectrum) at radio through sub-mm frequencies could be caused by factors such as adiabatic expansion losses, high magnetic fields ($>10^5$ G) at the base of the jet or a more confined jet geometry. Internal energy dissipations (e.g., internal shocks) in the jet can lead to multiple acceleration episodes as particles propagate along the jet \citep{kai06,jam10}, and thus could be an additional factor governing the observed spectral indices in this regime.
Further, noticeable inversion in the radio--sub-mm spectrum can occur if the jet plasma accelerates due to the longitudinal pressure gradient, as the Doppler factor changes non-linearly farther out from the central compact object  \citep{fal96,falmar,mar01}. The magnitude of such inversion is sensitive to many system parameters, in particular, the inclination of the system, where the inversion is more significant when the jet axis is not pointed along the line of sight (i.e., high inclinations). While this alone is not enough to explain the inversion in the Swift J1745$-$26 spectrum, it could be a contributing factor.  Additionally, the acceleration profile or the presence of additional, unaccounted for cooling processes could also be factors contributing to the spectral inversion.
  
Similar spectral behaviour can be seen in low luminosity AGN (LLAGN; \citealt{ho99}), such as Sgr A*, where the spectrum becomes more inverted when approaching sub-mm frequencies (i.e., the sub-mm bump; \citealt{zyl92,fal98,mel01,an05}). This sub-mm bump can be well explained by synchrotron emission \citep{falmar} originating in the region near the black hole from the base of a jet (although this sub-mm emission can also be explained by particles in the inner accretion flow; \citealt{yuan03}), which in turn suggests that similar conditions (e.g., geometry) may exist in the innermost regions of LLAGN and BHXRBs. We note that while Swift J1745-26 was orders of magnitude higher in $L/{L_{{\rm edd}}}$ than Sgr A * during this outburst, \cite{plot15} find that the differences in jet power between quiescent and hard state systems (leading to weaker particle acceleration and a cooler more compact jet base in quiescent systems) primarily leads to changes in the optically thin part of the spectrum rather then the optically thick part to which we are referring here. 

\subsection{Jet Spectral Break Frequency Lower Limit}

Based on the radio through optical measurements, we placed lower limits on the jet spectral break frequency of $\nu_{\rm break} \gtrsim 2\times10^{11}$ or $\nu_{\rm break} \gtrsim 1\times10^{14}\,{\rm Hz}$, depending on whether the optical emission arises from an accretion disc or the jet, respectively.
The former is consistent with jet spectral breaks in other BHXRBs, while the latter would be one of the highest lower limits placed on a jet spectral break in a BHXRB to date \citep[cf][]{rus12}. However, even the assumption that all the optical flux arises from jet emission does not require that Swift J1745$-$26 is an outlier among BHXRBs.

We observed Swift J1745$-$26 early in its outburst, only two to six days after its highest and hardest X-ray flux on 2012 September 18 \citep[MJD 56188.7;][]{bel12}, which corresponded to a 2--10 keV X-ray luminosity of $3\times10^{37} \frac{d}{3 {\rm \, kpc}} {\rm \, erg \, s^{-1}}$ \citep{mun13}. The discovery of a high frequency $\nu_{\rm break}$ near the peak hardness of an outburst would be consistent with an evolving jet break that evolves towards lower frequencies as X-ray hardness decreases (as seen in MAXI J1836--194; \citealt{rus13}; \citealt{rus14}). In addition, the discovery of a high frequency $\nu_{\rm break}$ near the peak luminosity (and thus the maximum accretion rate) of this outburst would be consistent with the \citet{rus14} results that suggest that evolution of $\nu_{\rm break}$ does not scale with the source luminosity in the hard state. 

However, the large uncertainty on the optical flux density measurements combined with the uncertainty in the optical emission processes in Swift J1745-26 prevents us from drawing any further conclusions within the scope of this paper.

\subsection{High mm/sub-mm Fluxes}
The few BHXRBs observed in the mm/sub-mm regime have been surprisingly bright when compared to radio frequency measurements (for example, XTE J1118$+$480; \citealt{fender01}, MAXI J1836$-$194; \citealt{rus13}, and Swift J1745$-$26).
Historically, astronomers have typically found flat type spectra in jetted sources \citep{fen06} and it has only been recently that more inverted spectra have been observed in some sources ($\alpha>0.2$; e.g., \citealt{rus13a}). As such, it has been suggested that these high fluxes could in fact be an anomalous spectral feature, as they do not fit in with the standard flat spectral picture. 
For example,  \cite{mar01} found an alternative model-fit for the SED of XTE J1118+480 (as opposed to \citealt{fender01} who fit the SED with a simple broken power-law), where the sub-mm (350GHz) flux is considered anomalous. 
Neglecting the sub-mm data point, the radio through X-ray emission can be almost entirely fit by synchrotron emission. As a result, \cite{mar01} find a flatter radio to IR spectral index than \cite{fender01}. This in turn results in a significant change in the location of the spectral break (from $\sim 40 \, \mu{\rm m}$ to $\sim 1 \, \mu{\rm m}$), implying different physical conditions in the jet. 

With recent evidence suggesting an evolving jet spectral break, we have to be careful when labelling high mm/sub-mm fluxes as anomalous if they are not compared to contemporaneous radio measurements (i.e., $\leq 1$ day). In XTE J1118$+$480, the sub-mm measurement was not simultaneous with the radio measurements \citep{fender01}, thus it is difficult to determine whether this measurement is in fact anomalous or not (as is the case with Cyg X--1; \citealt{fenpoo00}, and GRS 1915$+$105; \citealt{fenpoo002,og00}).
In MAXI J1836--194 \citep{rus13a}, the mm measurement is clearly consistent with the contemporaneous radio measurements, and the radio through mm spectrum is quite accurately represented by a single power-law through multiple epochs. Additionally, the Herschel detections of GX 339-4 at even higher frequencies ($70\,\mu {\rm m}$/$160\,\mu {\rm m}$ or $4282.7\,{\rm GHz}$/$1873.7\,{\rm GHz}$), were also consistent with the extrapolation of an inverted radio frequency power-law \citep{cor13}.

Similarly, in our global (radio through sub-mm) spectral fits, both the SMA (230GHz) mm and JCMT (350GHz) sub-mm measurements are consistent with contemporaneous radio measurements (with deviations $<1\sigma$). Therefore, we now have evidence from three sources that challenge the hypothesis that high mm/sub-mm fluxes are anomalous in all BHXRB sources, and supports our suggestion presented in the last section that the same mechanism driving spectral inversion, could be driving high mm/sub-mm fluxes. This result clearly justifies the need to continue to sample this mm/sub-mm regime more completely in multiple sources to confirm whether this result holds across the Galactic BHXRB population, especially considering our sparse temporal coverage in the mm/sub-mm regime and the high level of uncertainty in our sub-mm (350 GHz) measurement.

\subsection{Radio Frequency Variability}
In \S 3.1 and \S 3.2 we presented evidence for flux variability at radio frequencies that might be occurring on timescales of minutes (between bands within a VLA observation) to days (between VLA and SMA/JCMT observations).
The variability we see in the spectrum may result from the uncertainty introduced either in calibrating the data, interpolating the data, systematic errors introduced from combining data from different telescopes, or a combination of the three. However, previous observing campaigns of BHXRBs have revealed the presence of variability in the hard state (e.g., 
 GX 339--4 at radio through X-ray frequencies; \citealt{cor00,cor09}, and correlated IR-mm-radio flares in GRS 1915$+$105; \citealt{fenpoo002}).  
Further, jet models of \cite{mal13,mal13b,mal14} and \cite{jam10} suggest a possible mechanism that could lead to such rapid variability. In these models, collisions between discrete shells of plasma (injected at the base of the jet with variable bulk Lorentz factors) cause internal shocks in the jet that can naturally produce multi-wavelength variability, possibly occurring on minute timescales at radio frequencies. Such variation is smeared out in longer integrations but could cause the scatter we see here at radio frequencies. 
Conclusively distinguishing between these two options requires further in depth analysis of the data, and thus will be explored in future work.

\section{Summary}
In this paper, we have presented the results of our observations of the BHXRB source, Swift J1745$-$26 during its 2012--2013 outburst at radio and mm/sub-mm frequencies with the VLA, SMA and JCMT. This campaign marked both the first time that (quasi) simultaneous radio and multiple band mm \& sub-mm observations of a BHXRB have been obtained and the first time that the mm/sub-mm spectral index of a BHXRB jet has been measured. The combination of radio and mm/sub-mm measurements allowed us to compare the spectral behaviour between the two regimes and directly probe a part of the jet spectrum that has never been thoroughly sampled before. Through this work we aimed to test whether the jet emission we see was consistent with standard (power-law) jet models, as well as constrain the origin of the large mm/sub-mm fluxes we see in outbursting BHXRBs.

To analyze the jet spectrum in terms of standard jet models we fit a power-law model ($f_\nu \propto \nu^\alpha$, where $\alpha$ is constant) to the radio, mm/sub-mm, and global (radio through sub-mm) regimes for the different epochs when we had data. 

Our spectral fitting revealed a more inverted spectral index in data sets that contain mm/sub-mm measurements ($\alpha\sim0.07-0.17$), when compared to data sets containing radio-only measurements ($\alpha\sim0.03-0.05$).
Therefore, our measurements suggest a more inverted spectral index  
across radio--sub-mm frequencies is contributing to the high mm/sub-mm fluxes seen in BHXRBs, and perhaps the same mechanism behind the inversion may also be driving high mm/sub-mm fluxes seen in outbursting BHXRBs. Interestingly, this spectral inversion in the mm/sub-mm regime may be analogous to that seen in the sub-mm bump of LLAGN spectra (e.g., Sgr A*), possibly suggesting a common mechanism for bright sub-mm emission in LLAGN and BHXRBs.

Combining our radio and mm/sub-mm data with optical measurements from \cite{mun13}, we find that the i'band flux density may lie along the extrapolation of the radio--sub-mm power-law. Assuming that the i'band emission is entirely from the jet, this allows us to place 99\% confidence interval lower limits for the jet spectral break frequency of $\nu_{\rm break} \gtrsim 1\times10^{14}\,{\rm Hz}$ early in the outburst. While these are one of the highest lower limits placed on a jet spectral break in a BHXRB to date \citep[cf][]{rus12}, they are consistent with the emerging picture that links X-ray hardness and jet spectral break frequency, where a harder X-ray spectrum tends to have a higher $\nu_{\rm break}$ \citep{rus14}. However, as the optical flux density may contain significant contributions from an accretion disc, we can place an alternative lower limit to the jet spectral break frequency (using only radio and sub-mm data) of $2.3\times10^{11}\,{\rm Hz}$, with the optical flux density from the jet lying well below the extrapolated radio--sub-mm power-law. Such a lower limit to the jet spectral break frequency is similar to that seen in MAXI J1836$-$194.

While our data were reasonably well fit with a single power-law, small deviations at radio frequencies could suggest the possibility of rapid radio flux variability (timescales less than our observational cadence). 
However, these deviations from a single power-law are only known at $\sim 2\sigma$ confidence, and could be explained by poor quality data.
In addition to possible rapid variability, we find clear day to day variability at radio frequencies. Therefore, taking into account that we observe jet emission from the optical depth, $\tau=1$ surface at each frequency, resulting in an unknown travel time delay in the jet where variation will be observed at mm frequencies before cm frequencies, we conclude that obtaining simultaneous ($<1$ day) overlapping multi-wavelength observations across multiple epochs is necessary to accurately probe the jet spectrum in BHXRBs.

Although our results contain some uncertainty, they clearly point out the vital importance of the mm/sub-mm regime in understanding the jet spectrum, demonstrate the capacity of current mm/sub-mm instruments to address questions in this regime, and justify the need to explore this regime further. More high-quality, well-sampled SEDs of BHXRBs in outburst, including the mm/sub-mm regime, will help further constrain the jet spectrum, aid in developing more accurate jet models, and ultimately help understand the underlying physics of relativistic jets in BHXRBs.

\section{Acknowledgments}
AJT would like to thank Erik Rosolowsky for sharing his extensive knowledge of MCMC and Craig Heinke for helpful discussions. AJT and GRS are supported by an NSERC Discovery Grant. This work was supported by Australian Research Council grant DP120102393. This work was supported by the Spanish Ministerio de Econom\'ia y Competitividad and European Social Funds through a Ram\`on y Cajal Fellowship (S.M.) and the Spanish Ministerio de Ciencia e Innovaci\`on (S.M.; grant AYA2013-47447-C03-1-P). The Submillimeter Array is a joint project between the Smithsonian Astrophysical Observatory and the Academia Sinica Institute of Astronomy and Astrophysics, and is funded by the Smithsonian Institution and the Academia Sinica. 
The James Clerk Maxwell Telescope has historically been operated by the Joint Astronomy Centre on behalf of the Science and Technology Facilities Council of the United Kingdom, the National Research Council of Canada and the Netherlands Organisation for Scientific Research. Additional funds for the construction of SCUBA--2 were provided by the Canada Foundation for Innovation. The National Radio Astronomy Observatory is a facility of the National Science Foundation operated under cooperative agreement by Associated Universities, Inc.

\bibliography{referpaper}

\begin{thebibliography}{}
\expandafter\ifx\csname natexlab\endcsname\relax\def\natexlab#1{#1}\fi

\bibitem[{An {et~al.}(2005)An, Zhao, Hong, Roy, Rao, \& Shen}]{an05}
An, T., Zhao, J., Hong, X., {et~al.} 2005, ApJ, 634, L49

\bibitem[{Belloni(2010)}]{bel10}
Belloni, T. 2010, in The Jet Paradigm - From Microquasars to Quasars, ed.
  Belloni, T.M., Lecture Notes in Physics 794, Springer

\bibitem[{{Belloni} {et~al.}(2012){Belloni}, {Cadolle Bel}, {Casella},
  {Castro-Tirado}, {Corbel}, {Del Santo}, {Gallo}, {Grinberg}, {Homan},
  {Kalemci}, {Miller}, {Miller-Jones}, {Motta}, {Munoz-Darias}, {Nowak},
  {Pottschmidt}, {Rodriguez}, {Russell}, {Tomsick}, \& {Wilms}}]{bel12}
{Belloni}, T., {Cadolle Bel}, M., {Casella}, P., {et~al.} 2012, The
  Astronomer's Telegram, 4450, 1

\bibitem[{Blandford \& K\"onigl(1979)}]{blandford79}
Blandford, R., \& K\"onigl, A. 1979, ApJ, 232, 34

\bibitem[{Brocksopp {et~al.}(2004)Brocksopp, Bandyopadhyay, \& Fender}]{brok04}
Brocksopp, C., Bandyopadhyay, R., \& Fender, R. 2004, New Astronomy, 9, 249

\bibitem[{Cardelli {et~al.}(1989)Cardelli, Clayton, \& Mathis}]{card89}
Cardelli, J., Clayton, G., \& Mathis, J.~S. 1989, ApJ, 345, 245

\bibitem[{Casella \& Pe'er(2009)}]{cas09}
Casella, P., \& Pe'er, A. 2009, ApJ, 703, L63

\bibitem[{Casella {et~al.}(2010)Casella, Maccarone, O'Brien, Fender, Russell,
  van~der Klis, Pe'er, Maitra, Altamirano, Belloni, Kanbach, M., Mason, Soleri,
  Stefanescu, Wiersema, \& Wijnands}]{cas10}
Casella, P., Maccarone, T., O'Brien, K., {et~al.} 2010, MNRAS, 404, L21

\bibitem[{Chapin {et~al.}(2013)Chapin, Berry, Gibb, Jenness, Scott, Tilanus,
  Economou, \& Holland}]{chap}
Chapin, E., Berry, D., Gibb, A., {et~al.} 2013, MNRAS, 430, 2545

\bibitem[{Charles \& Coe(2006)}]{char06}
Charles, P.~A., \& Coe, M. 2006, in Compact Stellar X-ray Sources, ed. Lewin
  W.H.G., van der Klis M., Cambridge Astrophysics Series, No. 39, 215

\bibitem[{Chaty {et~al.}(2011)Chaty, Dubus, \& Raichoor}]{chat}
Chaty, S., Dubus, G., \& Raichoor, A. 2011, A\&A, 529, A3

\bibitem[{Corbel {et~al.}(2013{\natexlab{a}})Corbel, Coriat, Brocksopp,
  Tzioumis, Fender, Tomsick, M.M., \& Bailyn}]{cor12}
Corbel, S., Coriat, M., Brocksopp, C., {et~al.} 2013{\natexlab{a}}, MNRAS, 428,
  2500

\bibitem[{Corbel {et~al.}(2012)Corbel, Edwards, Tzioumis, Coriat, Fender, \&
  Brocksopp}]{cor12a}
Corbel, S., Edwards, P., Tzioumis, T., {et~al.} 2012, ATel, 4410, 1

\bibitem[{Corbel \& Fender(2002{\natexlab{a}})}]{corfen02}
Corbel, S., \& Fender, R. 2002{\natexlab{a}}, ApJ, 573, L35

\bibitem[{Corbel \& Fender(2002{\natexlab{b}})}]{cor02}
---. 2002{\natexlab{b}}, ApJ, 573, L35

\bibitem[{Corbel {et~al.}(2000)Corbel, Fender, Tzioumis, McIntyre, Durouchoux,
  \& Sood}]{cor00}
Corbel, S., Fender, R., Tzioumis, A.K. amd~Nowak, M., {et~al.} 2000, A\&A, 359,
  251

\bibitem[{Corbel {et~al.}(2013{\natexlab{b}})Corbel, Aussel, Broderick,
  Chanial, Coriat, Maury, Buxton, Tomsick, Tzioumis, Markoff, Rodriguez,
  Bailyn, Brocksopp, Fender, Petrucci, Cadolle-Bell, Calvelo, \&
  Harvey-Smith}]{cor13}
Corbel, S., Aussel, H., Broderick, J., {et~al.} 2013{\natexlab{b}}, MNRAS, 431,
  L107

\bibitem[{Coriat {et~al.}(2009)Coriat, Corbel, M.M., Bailyn, Tomsick, Kording,
  \& Kalemci}]{cor09}
Coriat, M., Corbel, S., M.M., B., {et~al.} 2009, MNRAS, 400, 123

\bibitem[{Cummings {et~al.}(2012{\natexlab{a}})Cummings, Gronwall, Grupe, H.A.,
  C.B., D.M., Sbarufatti, \& M.}]{cum12a}
Cummings, J., Gronwall, C., Grupe, D., {et~al.} 2012{\natexlab{a}}, GCN
  Circular, 13774, 1

\bibitem[{Cummings {et~al.}(2012{\natexlab{b}})Cummings, Barthelmy,
  Baumgartner, Fenimore, Gehrels, Krimm, Markwardt, Palmer, Sakamoto, Sato,
  Stamatikos, Tueller, \& Ukwatta}]{cum12b}
Cummings, J., Barthelmy, S., Baumgartner, W., {et~al.} 2012{\natexlab{b}}, GCN
  Circular, 13775, 1

\bibitem[{Curran {et~al.}(2014)Curran, Coriat, Miller-Jones, Armstrong,
  Edwards, Sivakoff, Woudt, Altamirano, Belloni, Corbel, Fender, Kording,
  Krimm, Markoff, Migliari, Russell, Stevens, \& Tzioumis}]{curr14}
Curran, P., Coriat, M., Miller-Jones, J., {et~al.} 2014, MNRAS, 437, 3265

\bibitem[{Dempsey {et~al.}(2012)Dempsey, Friberg, Jeness, Tilanus, Thomas,
  Holland, Bintley, Berry, Chapin, Chrysostomou, Davis, Gibb, Parsons, \&
  Robson}]{demp}
Dempsey, J., Friberg, P., Jeness, T., {et~al.} 2012, MNRAS, 430, 2534

\bibitem[{Fabian(2012)}]{fab12}
Fabian, A. 2012, ARA\&A, 50, 455

\bibitem[{Falcke(1996)}]{fal96}
Falcke, H. 1996, ApJ, 464, L67

\bibitem[{Falcke \& Biermann(1995)}]{falcke95}
Falcke, H., \& Biermann, P. 1995, A\&A, 293, 665

\bibitem[{Falcke {et~al.}(1998)Falcke, Goss, Matsuo, Teuben, Zhao, \&
  Zylka}]{fal98}
Falcke, H., Goss, W., Matsuo, H., {et~al.} 1998, ApJ, 499, 731

\bibitem[{Falcke {et~al.}(2004)Falcke, Kording, \& Markoff}]{falkormar04}
Falcke, H., Kording, E., \& Markoff, S. 2004, A\&A, 414, 895

\bibitem[{Falcke \& Markoff(2000)}]{falmar}
Falcke, H., \& Markoff, S. 2000, A\&A, 362, 113

\bibitem[{Fender(2001)}]{fen01}
Fender, R. 2001, MNRAS, 322, 31

\bibitem[{Fender(2006)}]{fen06}
---. 2006, in Compact Stellar X-ray Sources, ed. Lewin W.H.G., van der Klis M.,
  Cambridge Astrophysics Series, No. 39, 381

\bibitem[{Fender(2010)}]{fen10}
---. 2010, in The Jet Paradigm - From Microquasars to Quasars, ed. Belloni,
  T.M., Lecture Notes in Physics 794, Springer

\bibitem[{Fender \& Belloni(2004)}]{fenbel04}
Fender, R., \& Belloni, T. 2004, ARA\&A, 42, 317

\bibitem[{Fender {et~al.}(2004)Fender, Belloni, \& Gallo}]{fenbelgal04}
Fender, R., Belloni, T., \& Gallo, E. 2004, MNRAS, 355, 1105

\bibitem[{Fender {et~al.}(2001)Fender, Hjellming, Tilanus, Pooley, Deane,
  Ogley, \& Spencer}]{fender01}
Fender, R., Hjellming, R., Tilanus, R., {et~al.} 2001, MNRAS, 322, L23

\bibitem[{Fender {et~al.}(2009)Fender, Homan, \& Belloni}]{fenhombel09}
Fender, R., Homan, J., \& Belloni, T. 2009, MNRAS, 396, 1370

\bibitem[{Fender \& Pooley(2000)}]{fenpoo002}
Fender, R., \& Pooley, G. 2000, MNRAS, 318, L1

\bibitem[{Fender {et~al.}(2000)Fender, Pooley, Durouchoux, Tilanus, \&
  Brocksopp}]{fenpoo00}
Fender, R., Pooley, G., Durouchoux, P., Tilanus, R., \& Brocksopp, C. 2000,
  MNRAS, 312, 853

\bibitem[{{Foreman-Mackey} {et~al.}(2013){Foreman-Mackey}, {Hogg}, {Lang}, \&
  {Goodman}}]{for2013}
{Foreman-Mackey}, D., {Hogg}, D.~W., {Lang}, D., \& {Goodman}, J. 2013, \pasp,
  125, 306

\bibitem[{Gallo(2010)}]{gal10}
Gallo, E. 2010, in The Jet Paradigm - From Microquasars to Quasars, ed.
  Belloni, T.M., Lecture Notes in Physics 794, Springer

\bibitem[{Gallo {et~al.}(2005{\natexlab{a}})Gallo, Fender, \& Hynes}]{galc05}
Gallo, E., Fender, R., \& Hynes, R. 2005{\natexlab{a}}, MNRAS, 356, 1017

\bibitem[{Gallo {et~al.}(2005{\natexlab{b}})Gallo, Fender, Kaiser, Russell,
  Morganti, Oosterloo, \& Heinz}]{galfenkai05}
Gallo, E., Fender, R., Kaiser, C., {et~al.} 2005{\natexlab{b}}, Nature, 436,
  819

\bibitem[{Gallo {et~al.}(2007)Gallo, Migliari, Markoff, Tomsick, Bailyn, Berta,
  Fender, \& Miller-Jones}]{gala07}
Gallo, E., Migliari, S., Markoff, S., {et~al.} 2007, ApJ, 670, 600

\bibitem[{Gallo {et~al.}(2014)Gallo, Miller-Jones, Russell, Jonker, Homan,
  Plotkin, Markoff, Miller, Corbel, \& Fender}]{gal14}
Gallo, E., Miller-Jones, J., Russell, D., {et~al.} 2014, MNRAS, 445, 290

\bibitem[{Gandhi {et~al.}(2011)Gandhi, Blain, Russell, Casella, Malzac, Corbel,
  D'Avanzo, Lewis, Markoff, Cadolle~Bel, Goldoni, Wachter, Khangulyn, \&
  Mainzer}]{gan11}
Gandhi, P., Blain, A., Russell, D., {et~al.} 2011, ApJ, 740, L13

\bibitem[{Guver \& Ozel(2009)}]{guv09}
Guver, T., \& Ozel, F. 2009, MNRAS, 400, 2050

\bibitem[{Heinz \& Grimm(2005)}]{heigrimm}
Heinz, S., \& Grimm, H. 2005, ApJ, 633, 384

\bibitem[{Heinz \& Sunyaev(2003)}]{heisun03}
Heinz, S., \& Sunyaev, R. 2003, MNRAS, 343, L59

\bibitem[{Hjellming \& Johnson(1988)}]{hj88}
Hjellming, R., \& Johnson, K. 1988, ApJ, 328, 600

\bibitem[{Ho(1999)}]{ho99}
Ho, L. 1999, ApJ, 516, 672

\bibitem[{Holland {et~al.}(2013)}]{holl}
Holland, W., {et~al.} 2013, MNRAS, 430, 2513

\bibitem[{Hynes {et~al.}(2009)Hynes, Bradley, Rupen, Gallo, Fender, Casares, \&
  Zurita}]{hyn05}
Hynes, R., Bradley, C., Rupen, M., {et~al.} 2009, MNRAS, 399, 2239

\bibitem[{Jamil {et~al.}(2010)Jamil, Fender, \& Kaiser}]{jam10}
Jamil, O., Fender, R., \& Kaiser, C. 2010, MNRAS, 401, 394

\bibitem[{Kaiser(2006)}]{kai06}
Kaiser, C. 2006, MNRAS, 367, 1083

\bibitem[{Kalemci {et~al.}(2014)Kalemci, Arabaci, Guver, Russell, Tomsick,
  Wilms, Weidenspointner, Kuulkers, Falanga, Dincer, Drave, Belloni, Coriat,
  Lewis, \& Mu{\~n}oz-Darias}]{kale14}
Kalemci, E., Arabaci, M.~O., Guver, T., {et~al.} 2014, MNRAS, 445, 1288

\bibitem[{Malzac(2013{\natexlab{a}})}]{mal13}
Malzac, J. 2013{\natexlab{a}}, in The Innermost Regions of Relativistic Jets
  and Their Magnetic Fields, EPJ Web of Conferences, Volume 61 ed. J.L.
  G{\'o}mez

\bibitem[{Malzac(2013{\natexlab{b}})}]{mal13b}
---. 2013{\natexlab{b}}, MNRAS, 429, L20

\bibitem[{Malzac(2014)}]{mal14}
---. 2014, MNRAS, 443, 229

\bibitem[{Markoff {et~al.}(2001)Markoff, Falcke, \& Fender}]{mar01}
Markoff, S., Falcke, H., \& Fender, R. 2001, A\&A, 372, L25

\bibitem[{Markoff {et~al.}(2003)Markoff, Nowak, Corbel, Fender, \&
  Falcke}]{mar03}
Markoff, S., Nowak, M., Corbel, S., Fender, R., \& Falcke, H. 2003, A\&A, 397

\bibitem[{Markoff {et~al.}(2005)Markoff, Nowak, \& Wilms}]{marnowwil05}
Markoff, S., Nowak, M., \& Wilms, J. 2005, ApJ, 635, 1203

\bibitem[{McMullin {et~al.}(2007)McMullin, Waters, Schiebel, Young, \&
  Golap}]{mc07}
McMullin, J.~P., Waters, B., Schiebel, D., Young, W., \& Golap, K. 2007,
  Astronomical Data Analysis Software and Systems XVI, ed. R.A. Shaw, F. Hill
  and D.J. Bell, Astronomical Society of the Pacific Conference Series, Volume
  376, 127

\bibitem[{Meier(2001)}]{mei01}
Meier, D. 2001, ApJ, 548, L9

\bibitem[{Melia \& Falcke(2001)}]{mel01}
Melia, F., \& Falcke, H. 2001, ARA\&A, 39, 309

\bibitem[{Merloni {et~al.}(2003)Merloni, Heinz, \& Di~Matteo}]{merhezdi03}
Merloni, A., Heinz, S., \& Di~Matteo, T. 2003, MNRAS, 345, 1057

\bibitem[{Miller-Jones {et~al.}(2012)Miller-Jones, Sivakoff, \& on~behalf of
  the~larger JACPOT XRB~collaboration}]{mill12}
Miller-Jones, J., Sivakoff, G., \& on~behalf of the~larger JACPOT
  XRB~collaboration. 2012, ATel, 4394, 1

\bibitem[{Mirabel \& Rodriguez(1999)}]{mirrod99}
Mirabel, I., \& Rodriguez, L. 1999, ARA\&A, 37, 409

\bibitem[{Munoz-Darias {et~al.}(2013)Munoz-Darias, de~Ugarte~Postigo, Russell,
  Guziy, Gorosabel, Casares, Armas~Padilla, Charles, Fender, Belloni, Lewis,
  Motta, Castro-Tirado, Mundell, Sanchez-Ramirez, \& Thone}]{mun13}
Munoz-Darias, T., de~Ugarte~Postigo, A., Russell, D.~M., {et~al.} 2013, MNRAS,
  432, 1133

\bibitem[{Nowak {et~al.}(2005)Nowak, Wilms, Heinz, Pooley, Pottschmidt, \&
  Corbel}]{now05}
Nowak, M., Wilms, J., Heinz, S., {et~al.} 2005, ApJ, 626, 1006

\bibitem[{Ogley {et~al.}(2000)Ogley, Bell~Burnell, Fender, Pooley, \&
  Waltman}]{og00}
Ogley, R., Bell~Burnell, S., Fender, R., Pooley, G., \& Waltman, E. 2000,
  MNRAS, 317, 158

\bibitem[{Paredes {et~al.}(2000)Paredes, Marti, Peracaula, Pooley, \&
  Mirabel}]{pared00}
Paredes, J., Marti, J., Peracaula, M., Pooley, G., \& Mirabel, I. 2000, A\&A,
  357, 507

\bibitem[{Pe'er \& Casella(2009)}]{peer09}
Pe'er, A., \& Casella, P. 2009, ApJ, 699, 1919

\bibitem[{Pe'er \& Markoff(2012)}]{pmark}
Pe'er, A., \& Markoff, S. 2012, ApJ, 753, 177

\bibitem[{Plotkin {et~al.}(2015)Plotkin, Gallo, Markoff, Homan, Jonker,
  Miller-Jones, Russell, \& Drappeau}]{plot15}
Plotkin, R., Gallo, E., Markoff, S., {et~al.} 2015, MNRAS, 446, 4098

\bibitem[{Polko {et~al.}(2010)Polko, Meier, \& Markoff}]{pol10}
Polko, P., Meier, D., \& Markoff, S. 2010, ApJ, 723, 1343

\bibitem[{Polko {et~al.}(2013)Polko, Meier, \& Markoff}]{pol13}
---. 2013, MNRAS, 428, 587

\bibitem[{Polko {et~al.}(2014)Polko, Meier, \& Markoff}]{pol14}
---. 2014, MNRAS, 438, 559

\bibitem[{Rahoui {et~al.}(2011)Rahoui, Lee, Heinz, Hines, Pottschmidt, Wilms,
  \& Grinberg}]{rah11}
Rahoui, F., Lee, J., Heinz, S., {et~al.} 2011, ApJ, 736, 63

\bibitem[{Remillard \& McClintock(2006)}]{remmc06}
Remillard, R., \& McClintock, J. 2006, ARA\&A, 44, 49

\bibitem[{Russell {et~al.}(2006)Russell, Fender, Hynes, Brocksopp, Homan,
  Jonker, \& Buxton}]{rus06}
Russell, D., Fender, R., Hynes, R., {et~al.} 2006, MNRAS, 371, 1334

\bibitem[{Russell {et~al.}(2013{\natexlab{a}})Russell, Gallo, \&
  Fender}]{rus13}
Russell, D., Gallo, E., \& Fender, R. 2013{\natexlab{a}}, MNRAS, 431, 405

\bibitem[{Russell {et~al.}(2013{\natexlab{b}})Russell, Markoff, Casella,
  Cantrell, Chatterjee, Fender, Gallo, Gandhi, Homan, Maitra, Miller-Jones,
  O'Brien, \& Shahbaz}]{rus12}
Russell, D., Markoff, S., Casella, P., {et~al.} 2013{\natexlab{b}}, MNRAS, 429,
  815

\bibitem[{Russell {et~al.}(2013{\natexlab{c}})Russell, Russell, Miller-Jones,
  O'Brien, Soria, Sivakoff, Slaven-Blair, Lewis, Markoff, Homan, Altamirano,
  Curran, Rupen, Belloni, Cadolle~Bel, Casella, Corbel, Dhawan, Fender, Gallo,
  Gandhi, Heinz, Kording, Krimm, Maitra, Migliari, Remillard, Sarazin, Shahbaz,
  \& Tudose}]{rus13a}
Russell, D., Russell, T., Miller-Jones, J., {et~al.} 2013{\natexlab{c}}, ApJ,
  768, L35

\bibitem[{Russell {et~al.}(2014)Russell, Soria, Miller-Jones, Curran, Markoff,
  Russell, \& Sivakoff}]{rus14}
Russell, T., Soria, R., Miller-Jones, J., {et~al.} 2014, MNRAS, 439, 1390

\bibitem[{Sbarufatti {et~al.}(2013)Sbarufatti, Kennea, Stroh, Burrows, Evans,
  Beardmore, Krimm, \& Gehrels}]{sbar13}
Sbarufatti, B., Kennea, J., Stroh, M., {et~al.} 2013, ATel, 4782, 1

\bibitem[{Sbarufatti {et~al.}(2012)Sbarufatti, Kennea, Burrows, Campana,
  Gehrels, Markwardt, Cummings, Siegel, Krimm, \& Marshall}]{shar12}
Sbarufatti, B., B., Kennea, J., Burrows, D., {et~al.} 2012, ATel, 4383, 1

\bibitem[{van~der Horst {et~al.}(2013)van~der Horst, Curran, Miller-Jones,
  Linford, Gorosabel, Russell, de~Ugarte~Postigo, Lundgren, Taylor, Maitra,
  Guziy, Belloni, Kouveliotou, Jonker, Kamble, Paragi, Homan, Kuulkers, Granot,
  Altamirano, Buxton, M.M., Castro-Tirado, Fender, Garrett, Gehrels, Hartmann,
  Kennea, Krimm, Mangano, Ramirez-Ruiz, Romano, Wijers, R.A.M.J., \&
  Wijnands}]{van13}
van~der Horst, A., Curran, P., Miller-Jones, J., {et~al.} 2013, MNRAS, 436,
  2625

\bibitem[{van Paradijs \& McClintock(1995)}]{van98}
van Paradijs, J., \& McClintock, J. 1995, in X-Ray Binaries, ed. Lewin, W.H.G.,
  van Paradijs, J., van den Heuvel, E.P.J., Cambridge Astrophysics Series, No.
  26, 58

\bibitem[{Vovk {et~al.}(2012)Vovk, Ferrigno, Bozzo, Drave, Sanchez, Kuulkers,
  Bazzano, Del~Santo, Fiocchi, Natalucci, Tarana, Caballero, Goetz, Chenevez,
  den Hartog, Kuiper, \& Watanabe}]{vov12}
Vovk, I., Ferrigno, C., Bozzo, E., {et~al.} 2012, ATel, 4381, 1

\bibitem[{White {et~al.}(1995)White, Nagase, \& Parmar}]{white95}
White, N., Nagase, F., \& Parmar, A. 1995, in X-Ray Binaries, ed. Lewin,
  W.H.G., van Paradijs, J., van den Heuvel, E.P.J., Cambridge Astrophysics
  Series, No. 26, 1

\bibitem[{Yuan {et~al.}(2003)Yuan, Quataert, \& Narayan}]{yuan03}
Yuan, F., Quataert, E., \& Narayan, R. 2003, ApJ, 598, 301

\bibitem[{Zylka {et~al.}(1992)Zylka, Mezger, \& Lesch}]{zyl92}
Zylka, R., Mezger, P., \& Lesch, H. 1992, A\&A, 261, 119

\end{thebibliography}

\end{document}